\documentclass{llncs}
\usepackage{times}
\usepackage{graphicx,xspace}
\usepackage{tikz}
\usetikzlibrary{shapes}
\usetikzlibrary{fit}
\usetikzlibrary{shadows}
\usetikzlibrary{backgrounds}
\usetikzlibrary{arrows}

\usepackage{wrapfig}
\usepackage[hide]{ed}
\usepackage{lstomdoc}
\usepackage{paralist}
\def\MASally{{\small \textsf{MASally}}\xspace}
\def\omdoc{{\small \textsf{OMDoc}}\xspace}
\def\mmt{{\small \textsf{MMT}}\xspace}
\usepackage[utf8]{inputenc}
\usepackage{hetcasl}

\newcommand{\Hets}{\textsc{Hets}\xspace}

\title{Towards Ontological Support for Principle Solutions in Mechanical Engineering}

\author{
Thilo Breitsprecher \inst{1} \and 
Mihai Codescu \inst{2}\and 
Constantin Jucovschi \inst{3}\and
Michael Kohlhase \inst{3}\and
Lutz Schr\"oder \inst{2}\and
Sandro Wartzack\inst{1}}

\institute{Department of Mechanical Engineering,\\ Friedrich-Alexander-Universität Erlangen-N\"urnberg \and Department of Computer Science,
  FAU Erlangen-N\"urnberg \and Computer Science, Jacobs
  University Bremen }

\begin{document}

\maketitle

\begin{abstract}
  The engineering design process follows a series of standardized
  stages of development, which have many aspects in common with
  software engineering. Among these stages, the principle solution can
  be regarded as an analogue of the design specification, fixing as it
  does the way the final product works. It is usually constructed as
  an abstract sketch (hand-drawn or constructed with a CAD system)
  where the functional parts of the product are identified, and
  geometric and topological constraints are formulated.  Here, we
  outline a semantic approach where the principle solution is
  annotated with ontological assertions, thus making the intended
  requirements explicit and available for further machine processing;
  this includes the automated detection of design errors in the final
  CAD model, making additional use of a background ontology of
  engineering knowledge. We embed this approach into a
  document-oriented design workflow, in which the background ontology
  and semantic annotations in the documents are exploited to trace
  parts and requirements through the design process and across
  different applications.
\end{abstract}

\section{Introduction}

Much like software engineering design (in an ideal world), the design of artifacts in
mechanical engineering follows a multi-stage process in which abstract requirements are
successively refined into a final solution. In fact, this process of \emph{systematic
  engineering design} is to some degree standardized in models that bear substantial
resemblance to the V-model, such as the German VDI 2221~\cite{VDI2221}. However, only the
last stage in this process, corresponding to the actual implementation in software
engineering, has well-developed tool support, in the shape of CAD systems that serve to
document the final design. Other stages of the design process are typically documented in
natural language, diagrams, or drawings. There is little or no support available for
interconnecting the various stages of the design, let alone verifying that decisions made
in one stage are actually implemented in the next stage.

Here, we embark on a program to fill this gap, focusing for a start on
the last step in the development process, in which we are given a
\emph{principle solution} and need to implement this solution in the
final design, a CAD model. The principle solution fixes important
design decisions in particular regarding physical layout, materials,
and connections but does not normally carry a commitment to a fully
concrete physical shape. It is typically represented by a
comparatively simple drawing, produced using plain graphics programs
(e.g.\ within standard presentation tools) or even by hand. As such,
it has a number of interesting features regarding the way it does, and
also does not, convey certain information. The basic issue is that
while one does necessarily indicate only one concrete shape in the
drawing, not all aspects and details of this sketch are actually meant
to be reflected in the final design. Some of this is obvious; e.g.\ it
is clear that slight crinkles in a hand drawing are not intended to
become dents in the final product, and to some (possibly lesser)
degree it is also clear that not everything that is represented as a
straight line or a rectangle in a simple sketch will necessarily be
realized by the same simple geometry in the actual design. Other
aspects are less straightforward; e.g.\ symmetries in the drawing such
as parallelism of lines or equal lengths of certain parts, right
angles, and even the spatial arrangement and ordering of certain
components may constitute integral parts of the principle solution or
mere accidents of the sketch. Other aspects of the design may be
indicated by standard graphical symbolism; e.g.\ crosses often
represent bolts. To aid human understanding of the principle solution,
it is typically accompanied by a natural-language explanation that
(hopefully) clears up most of the ambiguities; other aspects of the
design are understandable only in the context of sufficient implicit
knowledge, i.e.\ based on the experience of the design engineer.

The approach we propose in order strengthen and explicate the links
between the stages of the design process is, then, to integrate the
documents associated to each stage into a unified document-oriented
workflow using a shared background ontology. This ontology should be
strong enough to not only record mere hierarchical terminologies but
also, in our concrete scenario of principle solutions, to capture as
far as possible the qualitative design intentions reflected in the
principle sketch as well as the requisite engineering knowledge
necessary for its understanding. Such an ontology will in particular
support the tracing of concepts and requirements throughout the
development process; we shall moreover demonstrate on an example how
it enables actual \emph{verification} of a final design against
constraints indicated in the principle solution.

Technically, we realize this approach by means of a modular semantic
middleware architecture, the \emph{Multi-Application Semantic Alliance
  Framework} (\MASally), which connects a system of knowledge
management web services to standard applications -- in particular
document players and CAD systems -- via a network of thin API handlers
that essentially make the framework parametric in the choice of CAD
system. Background knowledge and design intentions are represented in
a modular ontology that provides material for user assistance and
forms the basis for the verification of design constraints. The
formalized engineering knowledge required for the latter task is
managed within the heterogeneous logical framework provided by the
\emph{Heterogeneous tool set} \Hets~\cite{MossakowskiEA07},
with the \emph{Web Ontology Language (OWL)}~\cite{HorrocksEA03}
playing the role of the primary representation logic for the sake of
its good computational properties. Sources of ontological knowledge
include, besides manually extracted knowledge on engineering and basic
geometry, semantic annotations of the principle sketch and the
extraction of assertional knowledge from a CAD model. We illustrate
our framework by means of an example where we verify aspects of the
design of an assembly crane against the principle solution.

\section{A Document-Oriented Process with Background Knowledge}
\label{sec:process}
We recall the stages of the \emph{engineering design process} according
to VDI 2221~\cite{VDI2221}.

\begin{description}
\item\textbf{S1 Problem}: a concise formulation of the purpose of the product to be designed.
\item\textbf{S2 Requirements List}: a list of explicitly named properties of the envisioned
  product. It is developed in cooperation between designer and client and corresponds to
  the user specification document in the V-model.
\item\textbf{S3 Functional Structure}: a document that identifies the functional components
  of the envisioned product and puts them into relation with each other.
\item\textbf{S4 Principle Solution}: an abstract sketch capturing the most important aspects
  of the design.
\item\textbf{S5 Embodiment Design}: a CAD design that specifies the exact shape of the
  finished product.
\item\textbf{S6 Documentation}: accompanies all steps of the design process.
\end{description}
We will now drill in on step \textbf{S4}, since machine support, which
offers the most obvious handle for adding value using semantic
methods.

\subsection{Principle Solutions}\label{sub:principleSolution}
According to Pahl and Beitz~\cite{PahBei:ed07}, one can develop a
principle solution for a product by combining working principles that
correspond to the sub-functions identified in the function structure
of the product. The search for applicable working principles and their
ensuing combination in the principle solution is essential for the
further product development. For example, the manufacturing costs are
determined to a large extent by these decisions. However, a
combination of working principles cannot be fully evaluated until it
is turned into a suitable representation. At this highly creative
stage of the design process, the engineer does not want to consider
the formalities inherent to a full-fledged CAD system. For this
reason, probably the most common representations of principle
solutions in mechanical engineering are old-fashioned hand-drawn
sketches. Developing the principle solution mainly involves the
selection of materials, a rough dimensional layout, and other
technological issues. The design engineer can refer to various support
tools for support in the search for working principles, such as the
design catalogues of Roth \cite{Roth.1994} and Koller
\cite{Koller.1998}. The degree of detail of a sketch varies between
the two main levels of the design: while at the assembly level, the
focus is mainly on the topology of the product, to ensure
compatibility of the principles to be combined, at the level of parts
and sub-assemblies more attention is given to the actual shape of the
product to be developed. In the following, we discuss an example of a
representation of a principle solution.

\subsection{Case Study: An Assembly Crane}\label{sec:case-study}

\begin{wrapfigure}r{7cm}\vspace*{-1em}
 \includegraphics[scale=0.3]{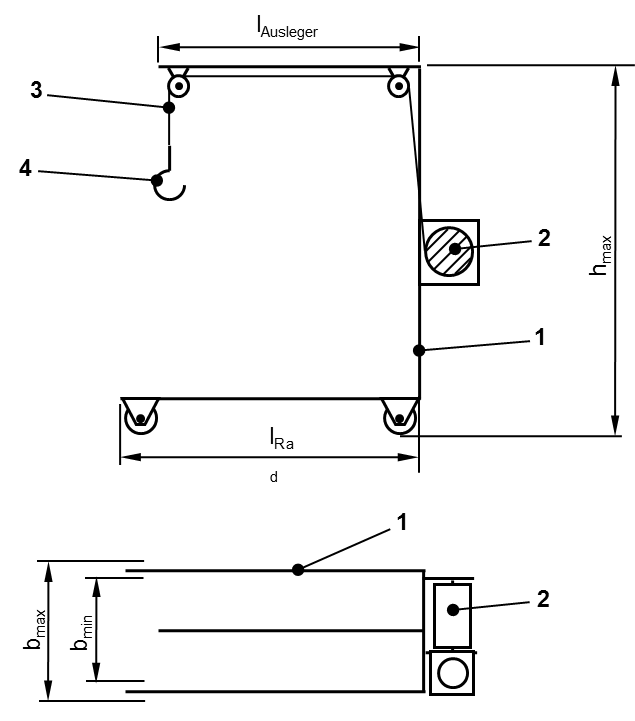}
 \caption{Principle Solution: Assembly Crane}\label{fig:crane}\vspace*{-2em}
\end{wrapfigure}

\begin{wrapfigure}l{7.3cm} \vspace*{-1em}
 \includegraphics[scale=.43]{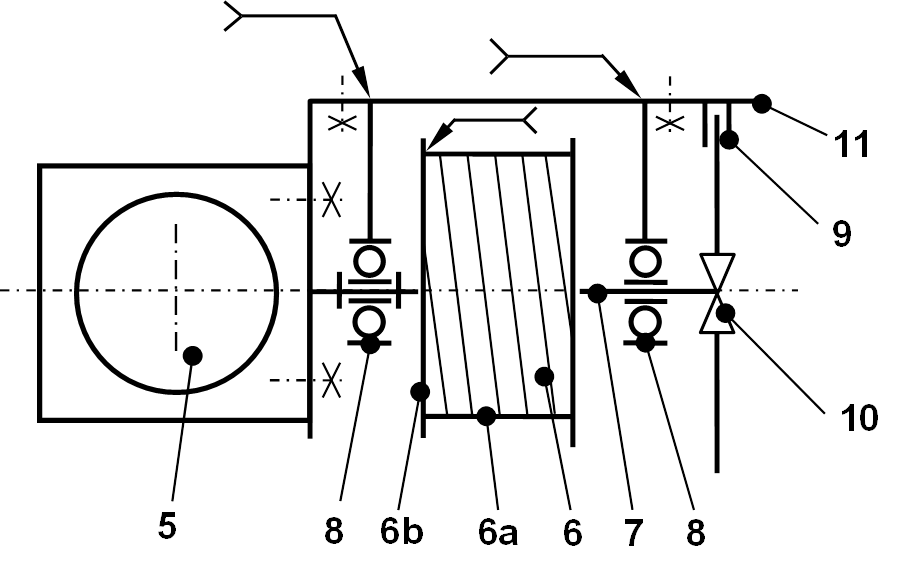}
 \caption{Principle Solution for the Winch Unit.}\label{fig:winch}\vspace*{-2em}
\end{wrapfigure}

Our main case study concerns an assembly crane for lifting heavy
machine components in workshops. This example has been used in a
practical design assignment for engineering students at the Chair of
Engineering Design at the University of Erlangen-Nürnberg in the
winter term of 2012. In this design exercise, students were given a
principle solution (Figures~\ref{fig:crane} and~\ref{fig:winch}) along
with some requirements (e.g.\ specified maximum power consumption,
maximum torque, and maximum weight) and were asked to design an
embodiment. Thus we have realistic documents for phases \textbf{S4}
and \textbf{S5} of a representative and non-trivial design task to
study.

The assembly crane to be designed can be divided into modules performing various
functions. The modules are indicated by numbers in the figure: the main frame with a
vertical beam, a cantilever, and parallel horizontal base profiles~(1); and a lifting
system, consisting of an electrically powered winch unit~(2), connected via a cable~(3),
which is guided via deflection rollers, to a crane hook~(4). This allows lifting, lowering
and holding the machine components to be assembled.  The requirements of the crane, which
have been defined in a previous step, concern the material to be used (standard steel
profiles for high strength and stiffness), the topology (the legs of the crane must be
parallel, the vertical and the horizontal cantilever are perpendicular, the motor (2) must
not be attached to the frame within the crane's working space), dimensions (maximum total
height, minimum space between base profiles, minimum cantilever length) and manufacturing
process constraints (weldment of main frame profiles and bolt connection of winch unit and
main frame).

Figure~\ref{fig:winch} details the principle solution of the winch unit: It consists of a
drum (6a), which is welded (generally, the requirement of a weldment is indicated
  by a folded arrow) between two side plates (6b). In order to ensure correct reeling of
the cable, the drum is thread-structured (6). The main shaft (7) is driven by an electric
worm-geared flange motor (5) that is connected to the winch frame (11) via blind-hole
bolts (indicated by crosses in the sketch). In order to decelerate the winch, to hold the
load, and to allow emergency stops, a lamella disk break (9) is installed; it is connected
to the main shaft by a suitable shaft-hub connection (10) that can withstand sudden
increases in torque (e.g.\ due to emergency stops). An arrangement of locating and
non-locating bearings (8) supports the main shaft. The ball bearings have to be arranged
in such a way that axial forces are kept from the motor. The winch frame is realized as a
stiff, yet weight-minimized, welded assembly, made of steel and is connected to the main
frame of the crane with through-hole bolts.

\section{Semantic Support for a Document-Oriented Engineering Workflow}\label{sec:docs}

Every step of the engineering design process results in particular documents, e.g.\ text
documents for \textbf{S1} to \textbf{S3} and \textbf{S6}, an image for \textbf{S4}
(hand-drawn or produced in a simple graphics program), and a CAD assembly in
\textbf{S5}. One of our goals is to integrate these into a document-oriented workflow,
using semantic technologies.

\subsection{A Semantic Annotation System}

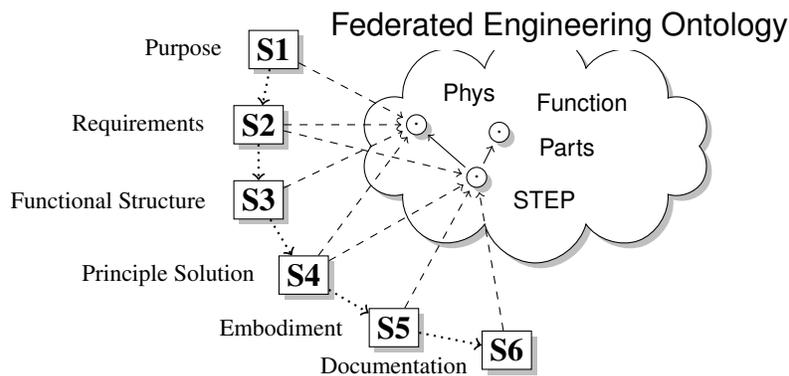
\begin{figure}[ht]\centering
\begin{tikzpicture}
  \tikzstyle{withshadow}=[draw,drop shadow={opacity=.5},fill=white]
  \tikzstyle{doc}=[withshadow]
  \tikzstyle{concept}=[circle,draw,inner sep=1pt,outer sep=2pt,withshadow]
  \tikzstyle{next}=[dotted,->,thick]
  \node at (-.8, 6) {\footnotesize Purpose};
  \node[doc] (s1) at (0.4,6) {\textbf{\large S1}};
  \node at (-1.4,5) {\footnotesize Requirements};
  \node[doc] (s2) at (0.2,5) {\textbf{\large S2}};
  \node at (-1.8,4) {\footnotesize Functional Structure};
  \node[doc] (s3) at (0.2,4) {\textbf{\large S3}};
  \node at (-1,3) {\footnotesize Principle Solution};
  \node[doc] (s4) at (0.8,3) {\textbf{\large S4}};
  \node at (0.5,2.3) {\footnotesize Embodiment};
  \node[doc] (s5) at (2,2.3) {\textbf{\large S5}};
  \node at (2,1.8) {\footnotesize Documentation};
  \node[doc] (s6) at (3.5,2) {\textbf{\large S6}};
  
  \draw[next] (s1) -- (s2);
  \draw[next] (s2) -- (s3);
  \draw[next] (s3) -- (s4);
  \draw[next] (s4) -- (s5);
  \draw[next] (s5) -- (s6);

  \node[concept] (in1) at (2.3,5) {$\cdot$};
  \node[concept] (in2) at (3.1,4.3) {$\cdot$};
  \node[concept] (in3) at (3.4,4.9) {$\cdot$};
  \draw[->] (in2) -- (in1);
  \draw[->] (in2) -- (in3);

  \node (step)  at (4,4) {\textsf{STEP}};
  \node (fun)  at (4.5,5.3) {\textsf{Function}};
  \node (parts) at (4.3,4.7) {\textsf{Parts}};
  \node (phys) at (3,5.4) {\textsf{Phys}};
  \begin{pgfonlayer}{background}
    \node[draw,cloud,fit=(step) (fun) (parts) (phys),aspect=2,inner sep=-5pt,withshadow] {};
  \end{pgfonlayer}
  \draw[->,dashed] (s1) -- (in1);
  \draw[->,dashed] (s2) -- (in1);
  \draw[->,dashed] (s3) -- (in1);
  \draw[->,dashed] (s4) -- (in1);

  \draw[->,dashed] (s4) -- (in2);
  \draw[->,dashed] (s5) -- (in2);
  \draw[->,dashed] (s6) -- (in2);
  \draw[->,dashed] (s2) -- (in2);

  \node[fill=white] (feo) at (4.2,6.3) {\textsf{\large Federated Engineering Ontology}};
\end{tikzpicture} 

\caption{An Ontology-Supported Document-Oriented Design Process}\label{fig:doconto}
\end{figure}

We build on the \MASally architecture presented
in~\cite{Kohlhase:kmsedcs13} (under the name {\small \textsf{FFCad}},
considerably extended here to embrace document-oriented workflows),
which assumes that the background knowledge shared by the
manufacturer, design engineer, and the clients is reified into a
flexiformal ontology (the cloud in Figure~\ref{fig:doconto}) and that
the documents are linked into that ontology via a semantic
illustration mapping. This illustration is a mapping (depicted by
dashed arrows in Figure~\ref{fig:doconto}) from fragments or objects
in the documents to concepts in the ontology (dotted circles), which
may themselves be interconnected by ontology relations (solid
arrows). The ontology (a term that we understand in a broad sense)
itself is a federation of ontology modules describing different
aspects of the engineering domain that are interconnected by
meaning-preserving interpretations (see Section~\ref{sec:feo} for
details).

\noindent In addition to the ontology links, we assume that the documents themselves are
semantically linked via relations (the dotted arrows between the \textbf{S$i$}) that model
the process of goal refinement in the development process. These two primary relations are
augmented with fine-grained annotations about document status, versioning, authorship,
etc. Note that our approach crucially extends metadata-based approaches in that the
annotations and relations point to document fragments -- e.g. text fragments down to
single symbols in formulae, regions in sketches, or shapes/sub-assemblies in CAD objects.

All of these explicit annotations in the documents are the basis for semantic services
that can be integrated into the documents (and their player applications) via the \MASally
framework, which we describe next.

\subsection{Semantic Services via the \MASally System}

The {\small \textsf{Multi-Application Semantic Alliance Framework}} (\MASally)
is a semantic middleware that allows embedding semantic interactions
into (semantically preloaded) documents. The aim of the system is to
support the ever more complex workflows of knowledge workers with
tasks that so far only other humans have been able to perform without
forcing them to leave their accustomed tool chain.

\begin{figure}[ht]\centering\vspace*{-1em}
\begin{tikzpicture}[node distance=0.8cm,scale=.9]\footnotesize
\tikzstyle{bigbox} = [draw=blue!50, thick, fill=blue!10, rounded corners, rectangle, minimum width=2.2cm]
\tikzstyle{box} = [minimum size=0.6cm, rounded corners,rectangle, fill=blue!50, font=\scriptsize]
\tikzstyle{database}=[cylinder, draw=blue!50, thick, shape border rotate=90, aspect=0.1, minimum height=6mm, minimum width=3cm,
       top color=blue!50,bottom color=blue!50,middle color=white,
       cylinder uses custom fill,
       cylinder end fill=white, font=\scriptsize, align=center] 
\tikzstyle{desktop} = [draw, thick, rounded corners, rectangle]
\tikzstyle{interaction} = [triangle 45-triangle 45, thick]

\node[box] (ass) {3D model {+ semlinks}};
\node[above of=ass,node distance=.6cm] (tcad) {CAD system};
\node[bigbox,below of=tcad, node distance=1.1cm] (alex) {\textsf{Alex}};
\begin{pgfonlayer}{background}
  \node[bigbox] [fit = (ass) (tcad) (alex)] (cad) {};
\end{pgfonlayer}
\node[bigbox, above of=cad, node distance=1.2cm] (theo) {\textsf{Theo}};

\node[box,below of=alex,node distance=2.3cm] (doc) {Text Model {+ semlinks}};
\node[above of=doc,node distance=.6cm] (tdoc) {Project Docs};
\node[bigbox,below of=tdoc, node distance=1.1cm] (dalex) {\textsf{Alex}};
\begin{pgfonlayer}{background}
  \node[bigbox] [fit = (doc) (tdoc) (dalex)] (tdoc) {};
\end{pgfonlayer}
\node[bigbox, above of=tdoc, node distance=1.2cm] (dtheo) {\textsf{Theo}};

\node[above of=theo,node distance=.7cm] (tdesktop) {Desktop};
\begin{pgfonlayer}{background}
  \node[desktop] [fit = (alex) (cad) (theo) (tdesktop) (dalex)] (desktop) {};
\end{pgfonlayer}

\node[box, right of=alex, node distance=3.7cm] (cmodel) {abs. CAD model};
\node[box, below of=cmodel,node distance=.7cm] (dmodel) {abs. text model};
\node[above of=cmodel] (tsally) {\textsf{Sally}};
\begin{pgfonlayer}{background}
\node[bigbox, fit = (dmodel) (tsally)] (sally) {};
\end{pgfonlayer}

\draw [interaction](sally) to node[above, right, yshift=2pt, xshift=-2pt, font=\tiny] {Comet} (theo.east);
\draw [interaction](sally) to node[above, right, yshift=2pt, xshift=-2pt, font=\tiny] {Comet} (dtheo.east);
\draw [interaction](cmodel) to node[below, xshift=5pt, font=\tiny] {Comet} (alex.east);
\draw [interaction](dmodel) to node[below, xshift=5pt, font=\tiny] {Comet} (dalex.east);

\node[database, right of=sally,node distance=4cm, yshift=1cm] (proj) {{Project
    documentation}\\{\em -- semiformal -- }}; 
\node[database, below of=proj, node distance=1.4cm] (background) {Background knowledge\\{(physics, engineering)}}; 
\node[database, below of=background, node distance=1.4cm] (iso) {{ISO/DIN norms}\\{\em --
    semiformal -- }}; 
\node[below of=iso, node distance=0.9cm] (dots) {$\vdots$};
\node[above of=proj, align=center, node distance=1cm] (ttnt) {\textsf{Planetary}};

\begin{pgfonlayer}{background}
\node[bigbox, fit = (proj) (background) (iso) (dots) (ttnt)] (tnt) {};
\end{pgfonlayer}
\draw [triangle 45-,thick](sally) to node[below, font=\tiny] {REST} (tnt);
\end{tikzpicture}
\vspace*{-1em}
\caption{The \MASally Architecture}\label{fig:masally}\vspace*{-1em}
\end{figure}
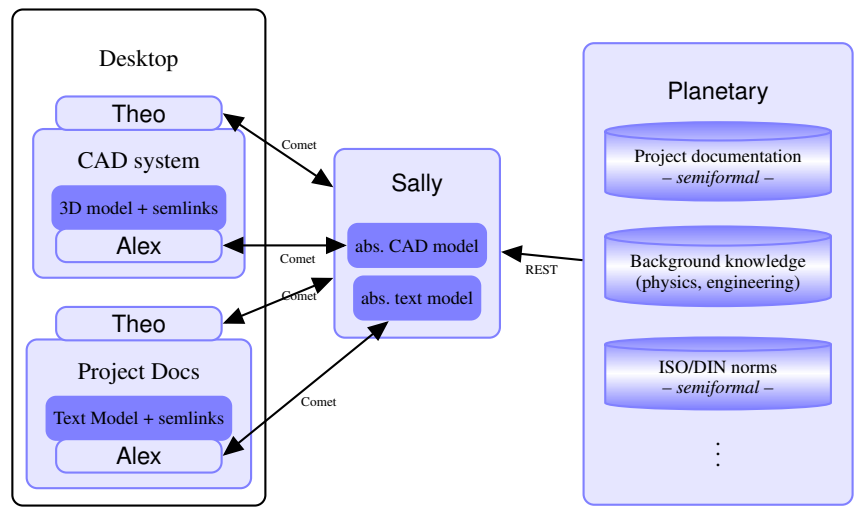
The \MASally system is realized as 
\begin{compactitem}
\item a set of semiformal knowledge management web services (comprised together with their
  knowledge sources under the heading {\small \textsf{Planetary}} on the right of
  Figure~\ref{fig:masally});
\item a central interaction manager ({\small \textsf{Sally}}, the \emph{semantic ally}) that
  coordinates the provisioning and choreographing of semantic services with the user
  actions in the various applications of her workflow;
\item and per application involved (we show a CAD system and a document viewer for
  \textbf{S4}/\textbf{S5} in Figure~\ref{fig:masally})
  \begin{compactitem}
  \item a thin API handler {\small \textsf{Alex}} that invades the application and relates its
    internal data model to the abstract, application-independent, content-oriented
    document model in {\small \textsf{Sally}};
  \item an application-independent display manager {\small \textsf{Theo}}, which super-imposes
    interaction and notification windows from {\small \textsf{Sally}} over the application window,
    creating the impression the semantic services they contain come from the
    application itself.
  \end{compactitem}
\end{compactitem}
This software and information architecture is engineered to share
semantic technologies across as many applications as possible,
minimizing the application-specific parts. The latter are encapsulated
in the {\small \textsf{Alex}}es, which only have to relate user events
to {\small \textsf{Sally}}, highlight fragments of semantic objects,
handle the storage of semantic annotations in the documents, and
export semantically relevant object properties to {\small
  \textsf{Sally}}. In particular, the {\small \textsf{Theo}}s are
completely system-independent. In our experience developing an {\small
  \textsf{Alex}} for an open-API application is a matter of less than
a month for an experienced programmer; see~\cite{DavJucKoh:safusa12}
for details on the \MASally architecture.

\begin{wrapfigure}r{6.8cm}\vspace*{-2em}
\includegraphics[width=6.8cm]{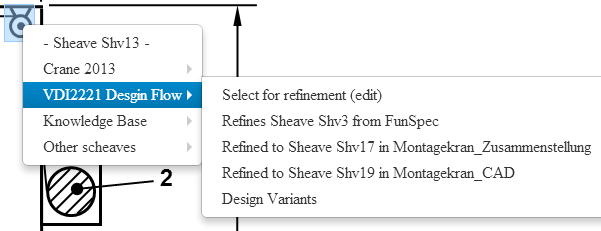}
\caption{Navigating the Refinement Relation}\label{fig:refinement}\vspace*{-1.5em}
\end{wrapfigure}
\begin{figure}[ht]\centering
  \includegraphics[width=9cm]{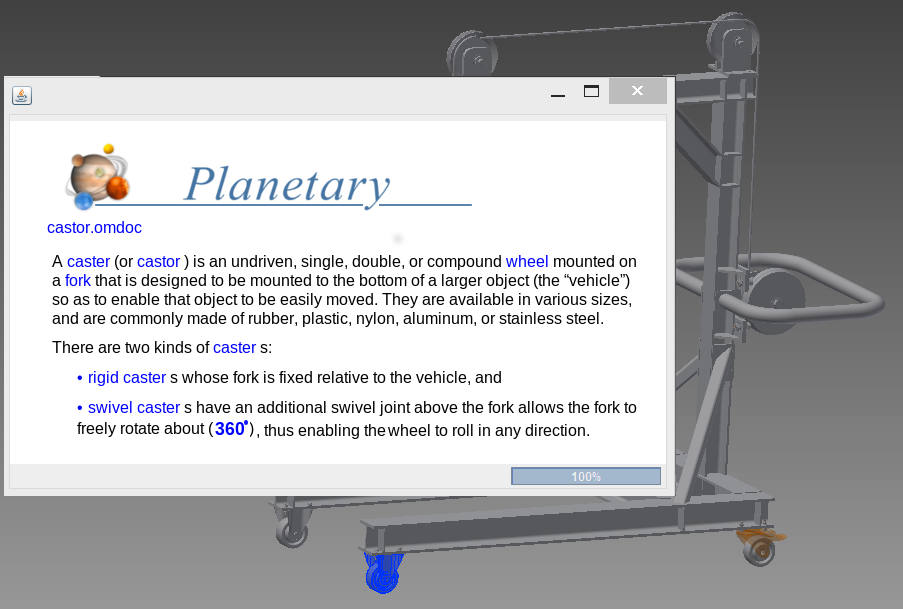}
  \caption{Definition Lookup}\label{fig:deflookup}
\end{figure}

To fortify our intuition about semantic services, let us consider the following
situation. The design engineer is working on the principle solution from Figure~\ref{fig:crane} -- a sketch
realized as a vector image, displayed in an (in this case browser-based) image viewer. The
user clicked on a detail of the sketch and received a {\small (\textsf{Theo}}-provided) menu
that
\begin{compactenum}
\item identifies the object as `Sheave S13' (the image is extended with an image map,
  which allows linking the region `S13' with the concept of a `sheave' in the
  ontology); further information about the object can be obtained by clicking
  on this menu item;
\item gives access to the project configuration that identifies the other documents in the
  current design;
\item gives access to the to the design refinement relation between the project documents:
  here, the object S13 is construed as a design refinement of the requirement S3 in the
  principle solution
  and has been further refined into objects S17 and S19 in the CAD
  assembly and the plans generated from that
  ;
\item allows direct interaction with the ontology (e.g. by definition lookup; see
  Figure~\ref{fig:deflookup}, here triggered from the CAD system for variety);
\item gives shortcuts for navigation to the other sheaves in the current project.
\end{compactenum}

Generally, the \MASally system supports generic help system functionalities (definition
lookup, exploration of the concept space, or semantic navigation: lookup of concrete CAD
objects from explanations) and allows focus-preserving task switching
(see~\cite{KohlhaseEtAl:FullSemanticTransparency:2013} for a discussion). All we need for
this are annotations of the VDI2221 relations, ontology links and of course the ontology
itself, which we will discuss next.

\section{The Federated Engineering Ontology}\label{sec:feo}
We now come to the design of the ontology that acts as the central
representation of the background knowledge and the common ground of
all actors in the design process.  It serves as a synchronization
point for semantic services, as a store for the properties of and
relations between domain objects, and as a repository of help texts
for the \MASally system. As it has to cover quite disparate
aspects of the respective engineering domain at different levels of
formality, it is unrealistic to expect a homogeneous ontology in a
single representation regime. Instead, we utilize the heterogeneous
\omdoc/\mmt framework~\cite{Kohlhase:OMDoc1.2,RabKoh:WSMSML13} that
allows representing and interrelating ontology modules via
meaning-preserving interpretations (i.e.\ theory morphisms). In
particular, \omdoc/\mmt supports the notion of meta-theories so that
we can have ontology modules represented in OWL2 alongside modules
written in higher-order logic, as well as informal modules given in
natural language. The \omdoc/\mmt meta-morphisms relate all of these
and moderate a joint frame of reference.  Reasoning support is
provided by the verification environment of the Heterogeneous Tool Set
\Hets~\cite{MossakowskiEA07}, a proof management tool that
interfaces state-of-the-art reasoners for logical languages.
\Hets mirrors the heterogeneity of the representation framework: new logics, logic
translations or concrete syntaxes of languages can be plugged in without having to modify
the heterogeneous and the deductive components of \Hets. In our verification of design
constraints, we employ, within \omdoc/\mmt/\Hets, the Distributed Ontology, Modeling and Specification Language
DOL \cite{MossakowskiEtAl13d,MossakowskiEA13} that provides specific support for heterogeneity in ontologies.

\subsection{A Verification Methodology}\label{sec:method}

We propose a general methodology for the verification of qualitative
properties of CAD assemblies against principle solutions. While the
checking of explicit \emph{quantitative} constraints in principle
solutions is supported by a number of research tools (e.g.\ the ProKon
system~\cite{Kratzer11}; in fact, some CAD systems themselves include
constraint languages such as CATIA Knowledge Expert, which however are
not typically interrelated with explicit principle solutions), there
is to our knowledge currently no support for checking
\emph{qualitative} requirements given by the principle solution.

\begin{example} \label{ex:tasks}In our case study introduced in
  Section~\ref{sec:case-study}, commonly encountered violations (in
  realizations produced by engineering students) of qualitative
  requirements in the principle solution were the following:
   \begin{itemize}
     \item the horizontal base profiles of the frame were not parallel;
     \item the types of weldments used did not ensure high stiffness and the local weldment area was not designed properly (e.g.\ missing ribs or stiffenings);
     \item the ball bearings were arranged in such a way that the non-locating bearing was closer to the motor, and thus the axial forces were transmitted into the motor.
   \end{itemize}
\end{example}
We are going to use the requirement that the legs of the frame should
be parallel as a running example throughout the rest of the
section. It is clear that the other examples can be treated similarly.

\begin{figure}[h]
\centering
\begin{tikzpicture}[every text node part/.style={align=center},every node/.style={drop shadow={opacity=.5},fill=white}]
  \node [rectangle, draw,rounded corners] (h) at (0,0) {Ontology of geometry};
  \node [rectangle, draw,rounded corners] (w) at (6,0) {Ontology of CAD features};
  \node [rectangle, draw,rounded corners] (e) at (3, -2) {Ontology of rules};
\draw (h) edge (e);
  \draw (w) edge (e);
  \node [rectangle, draw] (ps) at (0, -2.5) {\includegraphics[scale=0.1]{Bild1.png}};
  \node [rectangle, draw] (mod) at (6, -2.5) {\includegraphics[scale=0.15]{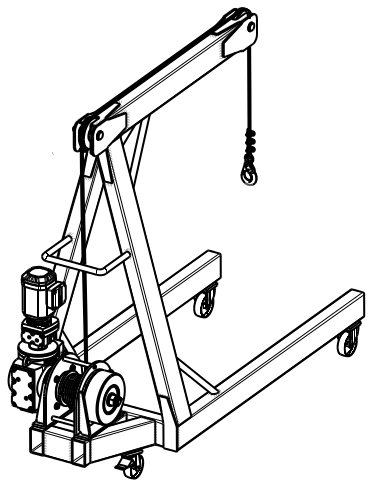}};
  \node [rectangle, draw] (sat) at (3, -4.5) { $\color{blue} T_M \textcolor{black}{\models} \textcolor{red}{T_R}$};
  \draw (ps) edge [color=red] (sat);
  \draw (mod) edge [color=blue](sat);
\end{tikzpicture}
\caption{Verification of qualitative properties of CAD designs.}\label{fig:methodology}
\end{figure}
\noindent The first step is to provide
a formal terminology for expressing the qualitative properties that a CAD design should fulfill. Since we are at the stage 
\textbf{S5} of the engineering design process, we have to collect requirements from all previous stages, in particular
\textbf{S1} - explicit requirements - and \textbf{S4} - further restrictions on the acceptable designs introduced by the
principle solution. Here, we  concentrate on geometric properties of physical objects and therefore we tackle this
goal by developing an ontology of geometric shapes.
We then need to have means to formally describe the aspects of a CAD design that are relevant for the properties that 
we want to verify. Since we want to verify geometric properties, we are going to make use of an ontology of
CAD features. We then need to formulate general rules regarding geometric properties of objects constructed by repeated 
applications of CAD features. This gives us a new ontology, of rules relating geometric properties and CAD features. 

We now come to the task of verification of a concrete CAD design against the requirements captured by a given principle 
solution. In a first step, we generate a representation of the requirements as an ABox $T_R$  over the ontology of rules, in a way that will be explained below. The next step is to generate a representation of the CAD design as another ABox $T_M$ over the same ontology of
rules, and then to make use of the rules to formally verify that $T_M$ logically implies $T_R$. This process is illustrated in 
Figure~\ref{fig:methodology}.

\subsection{Ontology of Shapes}

We begin setting up our verification framework by developing an
ontology of abstract geometric objects, with their shapes and
properties. The shape of a geometric object would seem to be a
well-understood concept; however, the task of formalizing the
semantics of shapes and reasoning about them is difficult to achieve
in a comprehensive way. For a broader discussion, including some
attempts to develop ontologies of geometric shapes, see, e.g., the
proceedings of the Shapes workshop~\cite{Shapes}. 


Our ontology, inspired by CYC \cite{cyc}, concentrates on geometric primitives of interest for 
CAD design . The central concept
is that of \Id{PhysicalObject}, which may be of an unspecified shape
or can have a 2-dimensional or 3-dimensional shape. Moreover, a \Id{ PhysicalObject} may be rigid or mobile, and holes are
represented as \Id{NegativeShapedThing}s. The known shapes, organized in a taxonomy, provide further concepts.
The object and data properties of the ontology are either parameters of the geometric shapes
(e.g.~diameter of a circle, or length of the sides of a square) or general geometric properties, like
symmetric 2D- and 3D-objects and parallel lines.

\begin{example}\label{ex:geom-onto}

We present the fragment of the ontology of shapes that is relevant for asserting that two objects are parallel.
This is given as a DOL specification that extends our OWL formalization of geometry with the
axiom that two lines are parallel if the angles of their intersections with a third line are equal.
Since the intersection of two lines is a relation with three arguments, the two intersecting lines and the angle formed by them,
we use reification to represent it as a concept \Id{Intersection},
together with a role  \Id{intersectsWith} that gives for an intersection the first constituent line,
a class \Id{LineAngle} for pairs of lines with angles (together with projection relations
\Id{hasAngle} and \Id{hasLine})
and a role \Id{lineAngleOf} giving the pair of the second line of an intersection with the angle between the two lines.
We use Manchester syntax for OWL, with {\bf o} denoting role composition. 
 
 \begin{hetcasl}
\SPEC \=\SIdIndex{Geom} \Ax{=}$\ldots$\\
\KW{ObjectProperty:} \=\Id{isParallelWith}\\
\KW{Domain:} \Id{PhysicalObject}\\
\KW{Range:} \Id{PhysicalObject}\\
\KW{SubPropertyChain:} \=\Id{hasIntersection} \KW{o} \Id{hasLineAngle} \KW{o} \Id{lineAngleOf} \KW{o} \Id{intersectsWith}
\end{hetcasl}

\end{example}

\subsection{Ontology of CAD Features}

Inspired by \cite{journals/ijcat/BrunettiG05}, our ontology of features
contains information about the geometry and topology of the CAD parts. 
Its concepts are assemblies and their parts, feature constructors and transformers, 2D sketches and their
primitives, or constraints (see Example~\ref{ex:feat-onto} below). 
Object and data properties are introduced for parameters of primitives, binary constraints or 
for composition rules (an assembly is formed with parts, a part has a 2D sketch base etc.).

\begin{example}\label{ex:feat-onto}

We present again in detail only a fragment of the ontology of features that is relevant for
verifying that two objects are parallel. We have a concept of \Id{Part} of an assembly and
each part has been constructed in a 3D space which has 3 axes of reference. We record this
by an object property \Id{hasAxis}, with the inverse \Id{isAxisOf}. Furthermore, 3D parts can be \emph{constrained}
at the assembly level. The constraint of interest for us is an angle constraint that specifies the angle 
formed between two axes, two edges or two faces of two chosen parts. Since this is again a relation with
three arguments, we use reification, in a similar way as in Example~\ref{ex:geom-onto}, that is, 
we have a class \Id{AngleConstraint} and three roles, \Id{fstLine} and \Id{sndLine} giving the two lines that are
constrained and \Id{angle} giving the specified angle.
  

\end{example} 

\subsection{Ontology of rules}

The next step is to relate via rules the concrete designs using 
feature transformers and constructors, given as elements of the ontology of features,
to the abstract shapes in the ontology of geometry. 
It is worth mentioning that the rules can be themselves subject to verification (a proof of concept was given in~\cite{KohLemSchSch:fmccp09a}). 
The advantage of our approach is that the task of verifying the rules, which can be quite complex, is separated from the task of 
checking correctness of individual CAD designs, which makes use of the rules. 

\begin{example}\label{ex:onto-rules}

We record below that each part is a geometric object, that an angle constraint in an assembly gives rise to an
intersection between the constrained lines and that two parts of an assembly are parallel if their axes are parallel.

\begin{hetcasl}
\SPEC \=\SIdIndex{Rules} \Ax{=} \SId{Features} \AND \SId{Geom}\\
\THEN \=\KW{Class:} \=\Id{Part}\\
\>\KW{SubClassOf:} \Id{PhysicalObject}\\
\> \KW{ObjectProperty:} \=\Id{isParallelWith}\\
\> \KW{SubPropertyChain:} \=\Id{hasAxis} \KW{o} \Id{isParallelWith} \KW{o} \Id{isAxisOf}\\
\> \KW{ObjectProperty:} \=\Id{fstLine}\\
\> \KW{SubPropertyChain:} \=\Id{intersectionOfAngle} \KW{o} \Id{intersectsWith}\\
\> \KW{ObjectProperty:} \=\Id{sndLine}\\
\> \KW{SubPropertyChain:} \=\Id{intersectionOfAngle} \KW{o} \Id{hasLineAngle} \KW{o} \Id{hasLine}\\
\> \KW{ObjectProperty:} \=\Id{angle}\\
\> \KW{SubPropertyChain:} \=\Id{intersectionOfAngle} \KW{o} \Id{hasLineAngle} \KW{o} \Id{hasAngle}
\end{hetcasl}

\end{example}

\subsection{Generating the ABoxes and proving correctness}

\begin{figure}[ht] \centering
 \includegraphics[width=11cm]{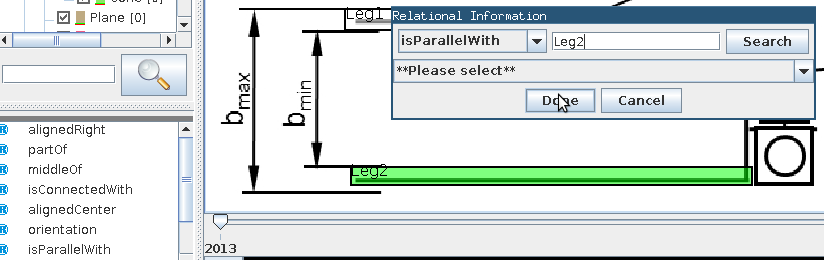}
 \caption{
 Making assertions regarding individuals explicit
    using AKTiveMedia \cite{chakravarthy2006aktivemedia}}\label{fig:annot_principle_sol}
\end{figure}

The principle solution is available as an image file, together with a
text document that records additional requirements introduced in the
principle solution, thus further restricting the acceptable
realizations of the design. Each part of the sketch has been
identified as a functional part of the principle solution and given a
name; this yields the required individual names for our ABox. The
assertions regarding the individuals thus obtained are added as
semantic annotations to the text that accompanies the image (Figure
\ref{fig:annot_principle_sol}).

\begin{example}\label{ex:geom-abox}
The following ABox expresses that the parts identified as \Id{leg1} and \Id{leg2} of the principle solution
should be parallel: 
\begin{hetcasl}
\SPEC \=\SIdIndex{PS\_ABox} \Ax{=} \SId{Geom}\\
\THEN \=\KW{Individual:} \=\Id{leg2}\\
\> \KW{Individual:} \=\Id{leg1}\\
\> \KW{Facts:} \Id{isParallelWith} \Id{leg2}
\end{hetcasl}
The ABox of the CAD design is generated from its history of
construction, using the Alex for CAD.  The following part of this ABox
expresses that the two legs of the crane have been explicitly
constrained to be perpendicular to the main frame in the CAD model:

\begin{hetcasl}
\SPEC \=\SIdIndex{CAD\_ABox} \Ax{=} \SId{Features}\\
\THEN \=\KW{Individual:} \=\Id{a1} \KW{Types:} \Id{Line}\\
 \KW{Individual:} \=\Id{a2} \KW{Types:} \Id{Line}\\
 \KW{Individual:} \=\Id{a3} \KW{Types:} \Id{Line}\\
 \KW{Individual:} \=\Id{leg1}  \KW{Types:} \Id{Part} \KW{Facts:} \Id{hasAxis} \Id{a1}\\
 \KW{Individual:} \=\Id{leg2} \KW{Types:} \Id{Part} \KW{Facts:} \Id{hasAxis} \Id{a2}\\
 \KW{Individual:} \=\Id{frameBase} \KW{Types:} \Id{Part} \KW{Facts:} \Id{hasAxis} \Id{a3}\\
 \KW{Individual:} \=\Id{alpha} \KW{Types:} \Id{Angle} \KW{Facts:} \Id{valueOf} 90\\
 \KW{Individual:} \=\Id{ac1} \KW{Types:} \Id{AngleConstraint}\KW{Facts:} \=\Id{fstLine} \Id{a1}, \Id{sndLine} \Id{a3}, \Id{angle} \Id{alpha}\\
 \KW{Individual:} \=\Id{ac2} \KW{Types:} \Id{AngleConstraint} \KW{Facts:} \=\Id{fstLine} \Id{a2}, \Id{sndLine} \Id{a3}, \Id{angle} \Id{alpha}
\end{hetcasl}
\end{example}
To complete all gaps in Figure~\ref{fig:methodology}, we have to show that all models (in the sense of interpretations of a logical theory) of the ABox generated from the CAD design are models of the 
ABox generated from the principle solution. DOL uses \emph{views} to express that, as in the example below.

\begin{example}
Checking that the two legs of the crane are parallel amounts to checking correctness of the  DOL view
\begin{hetcasl}
\VIEW \=\SId{Verif} \Ax{:} \SId{PS\_ABox} \KW{to} \SId{CAD\_ABox}
\end{hetcasl}

\noindent using one of the provers interfaced by \Hets, e.g.~ the Pellet reasoner for OWL \cite{pellet}; as expected, the reasoner makes short work of this.

\end{example}

\section{Related Work}

In previous work~\cite{KohLemSchSch:fmccp09a}, we have developed an
export function from {\sc SolidWorks} that generates from the internal
representation of a CAD design a description of its construction in a
variant of higher-order logic. One can then relate this construction
to abstract geometric shapes and prove this relation to be correct
using a higher-order proof assistant. In the context of the
methodology introduced in Section~\ref{sec:method}, each such relation
between the construction and its abstract geometric counterpart
gives rise to a formally verified rule in the ontology of rules. At
the informal level, we have moreover developed a semantic help system
for CAD objects based on the Semantic Alliance
Framework~\cite{Kohlhase:kmsedcs13}, and have illustrated the use of
this information for semantically supported task
switching~\cite{KohlhaseEtAl:FullSemanticTransparency:2013}.

Several ontologies of features have been developed, with the typical scenario being interoperability and data interchange between 
CAD systems, rather than verification of qualitative properties of CAD assemblies. 
We mention here only OntoSTEP \cite{journals/cad/BarbauKSNFFS12}, which aims to enrich the semantics of CAD objects
when exported using the ISO-standard interchange format STEP; it has the advantage of being independent of the choice of CAD system.
Our heterogeneous approach allows integrating OntoSTEP (or any other ontology
of features) into our federated engineering ontology and relating it to our ontology of features, without
having to modify our verification methodology.

Various approaches have been explored to integrate semantics into the
engineering design process. One such approach is the so-called
\emph{feature technology}, which has been researched by several
institutes. According to \cite{VDI2218}, features are an aggregation
of geometry items and semantics. Different types of features are
defined (eg. form features, semantic features, application features,
compound features), depending strongly on the technical domain and the
product life-cycle phase in which features are used. We expect
features to play a role in further semanticizing step S5 (embodiment,
Section~\ref{sec:process}) in future work.

The computer-supported tracing of the design relations described in Section~\ref{sec:docs}
is related to techniques of requirements tracing (RT; see~\cite{INCOSE} for an overview
over the current state of the art).  Current RT approaches are restricted to software
engineering workflows; they are usually dissociated from the documents that describe the
(software) artifacts and manage the requirements separately. We contend that our approach
that puts the VDI 2221 documents into the center of the workflow  integrates more directly
with existing engineering design workflows

As mentioned in Section \ref{sub:principleSolution}, the most common
representation of a principle solution in mechanical engineering are
probably old-fashioned hand-drawn sketches. However, alternative
approaches have been developed. Albers \cite{Albers13} proposes the
general Contact-and-Channel Model (CCM) for a model-based engineering
design process. The basic idea is that every technical system can be
represented as a system of \textit{working surface pairs} and
\textit{channel and support structures}. In our case study, an example
of a working surface pair would be the shaft-hub connection between
the winch main shaft and the lamella disk break, and the main shaft,
where the break torque of the disk break is \textit{channeled} to the
winch drum in order to stop the cable, would be a channel and support
structure. Approaches of this kind are candidates for integration
with our ontological process model in future extensions covering the
step from the function structure to the principle solution.

\section{Conclusions}

We have described a framework for semantic support in engineering
design processes, focusing on the step from the principle solution to
the embodiment, i.e.\ the CAD model. We base our framework on a
flexiformal background ontology that combines informal and semiformal
parts serving informational purposes with fully formalized qualitative
engineering knowledge and support for annotation of principle sketches
with formal qualitative constraints. The latter serve to separate
contingencies of the sketch from its intended information content, and
enable \emph{automated} verification of the CAD model against aspects
of the principle solution. We combine this approach with a
document-oriented workflow that also relies on the background ontology
for tracking the identity of parts through the design process and
across different applications, which are accessed in a unified manner
within the \MASally framework.

We have illustrated our approach on the partial verification of a CAD
model of an assembly crane, showing in particular that the ability to
draw logical inferences is important when verifying qualitative
constraints (related pre-existing systems support only quantitative
constraints, typically verified by direct calculation). The
logic-based approach thus allowed the system to, e.g., accept two
parts as satisfying a parallelism constraint formulated in the
principle solution although the CAD model did not directly contain
such a constraint, which instead had to inferred by combining other
constraints in the model.

\ednote{mention that the ABox extraction from CAD is currently under
  development?}  We currently use OWL as the logical core of our
verification framework, representing the requisite background
knowledge in a TBox and generating ABoxes from the principle sketch
and the CAD model. In principle, our approach is logic-agnostic, being
based on heterogeneous principles, in particular through use of the
Heterogeneous Tool Set \Hets and the Distributed Ontology, Modeling
and Specification Language DOL \cite{MossakowskiEtAl13d,MossakowskiEA13}. It is thus
easily possible to go beyond the expressivity boundaries of OWL where
necessary, e.g.\ by moving some parts of (!) the ontology into
first-order logic or, more conservatively, by using rule-based
extensions of OWL such as SWRL~\cite{HorrocksEA05} --- this will
increase the complexity of reasoning but the \Hets system will
localize this effect to those parts of the ontology that actually
need the higher expressive power. Use of SWRL will in particular
increase the capabilities of the system w.r.t.\ arithmetic reasoning.

\textbf{Acknowledgements.} We thank Oliver Kutz and Till Mossakowski for fruitful
discussions. The work presented in this paper was supported by the German Research
Foundation (DFG) under grant KO-2484/12-1 / SCHR-1118/7-1 (FormalCAD).

\bibliographystyle{myabbrv}

\begin{thebibliography}{10}

\bibitem{Albers13}
A.~Albers and C.~Zingel.
\newblock Extending {S}ys{ML} for engineering designers by integration of the
  contact and channel–approach ({CCM}) for function-based modeling of
  technical systems.
\newblock In {\em Systems Engineering Research, CSER 2013}, vol.~16 of {\em
  Proc.\ Comput.\ Sci.}, pp. 353 -- 362. Elsevier, 2013.

\bibitem{journals/cad/BarbauKSNFFS12}
R.~Barbau, S.~Krima, R.~Sudarsan, A.~Narayanan, X.~Fiorentini, S.~Foufou, and
  R.~D. Sriram.
\newblock {OntoSTEP}: Enriching product model data using ontologies.
\newblock {\em Computer-Aided Design}, 44:575--590, 2012.

\bibitem{journals/ijcat/BrunettiG05}
G.~Brunetti and S.~Grimm.
\newblock Feature ontologies for the explicit representation of shape
  semantics.
\newblock {\em J.\ Comput.\ Appl.\ Technology}, 23:192--202, 2005.

\bibitem{chakravarthy2006aktivemedia}
A.~Chakravarthy, F.~Ciravegna, and V.~Lanfranchi.
\newblock Aktivemedia: Cross-media document annotation and enrichment.
\newblock In {\em Semantic Web Annotation of Multimedia (SWAMM-06)}, 2006.

\bibitem{DavJucKoh:safusa12}
C.~David, C.~Jucovschi, A.~Kohlhase, and M.~Kohlhase.
\newblock \texttt{Semantic Alliance}: A framework for semantic allies.
\newblock In J.~Jeuring, J.~A. Campbell, J.~Carette, G.~Dos~Reis, P.~Sojka,
  M.~Wenzel, and V.~Sorge, eds., {\em {Intelligent Computer Mathematics}},
  number 7362 in LNAI, pp. 49--64. Springer Verlag, 2012.

\bibitem{cyc}
D.Lenat.
\newblock {Cyc: A Large-Scale Investment in Knowledge Infrastructure}.
\newblock {\em Communications of the ACM}, 38(11):33--38, 1995.

\bibitem{HorrocksEA05}
I.~Horrocks, P.~Patel-Schneider, S.~Bechhofer, and D.~Tsarkov.
\newblock {OWL} rules: A proposal and prototype implementation.
\newblock {\em J.\ Web Sem.}, 3:23--40, 2005.

\bibitem{HorrocksEA03}
I.~Horrocks, P.~F. Patel-Schneider, and F.~van Harmelen.
\newblock From {SHIQ} and {RDF} to {OWL}: the making of a web ontology
  language.
\newblock {\em Journal of Web Semantics}, 1(1):7--26, 2003.

\bibitem{INCOSE}
International council on systems engineering website.
\newblock \url{http://www.incose.org}.

\bibitem{KohlhaseEtAl:FullSemanticTransparency:2013}
A.~Kohlhase, M.~Kohlhase, C.~Jucovschi, and A.~Toader.
\newblock Full semantic transparency: Overcoming boundaries of applications.
\newblock In P.~Kotz\'{e}, G.~Marsden, G.~Lindgaard, J.~Wesson, and
  M.~Winckler, eds., {\em Human-Computer Interaction -- INTERACT 2013}, vol.
  8119 of {\em Lecture Notes in Computer Science}, pp. 406--423. Springer,
  2013.

\bibitem{Kohlhase:OMDoc1.2}
M.~Kohlhase.
\newblock {\em \textsc{OMDoc} -- An open markup format for mathematical
  documents [Version 1.2]}.
\newblock Number 4180 in LNAI. Springer Verlag, Aug. 2006.

\bibitem{Kohlhase:kmsedcs13}
M.~Kohlhase.
\newblock Knowledge management for systematic engineering design in {CAD}
  systems.
\newblock In F.~Lehner, N.~Amende, and N.~Fteimi, eds., {\em Professionelles
  Wissenmanagement Management, Konferenzbeitr{\"a}ge der 7. Konferenz}, pp.
  202--217. GITO Verlag, 2013.

\bibitem{KohLemSchSch:fmccp09a}
M.~Kohlhase, J.~Lemburg, L.~Schr{\"o}der, and E.~Schulz.
\newblock Formal management of {CAD/CAM} processes.
\newblock In A.~Cavalcanti and D.~Dams, eds., {\em 16\textsuperscript{th}
  International Symposium on Formal Methods (FM 2009)}, number 5850 in LNCS,
  pp. 223--238. Springer Verlag, 2009.

\bibitem{Koller.1998}
R.~Koller and N.~Kastrup.
\newblock {\em {P}rinziplösungen zur Konstruktion technischer Produkte}.
\newblock Springer, 1994.

\bibitem{Kratzer11}
M.~Kratzer, M.~Rauscher, H.~Binz, and P.~G{\"o}hner.
\newblock Konzept eines {W}issensintegrationssystems zur benutzerfreundlichen,
  benutzerspezifischen und selbst{\"a}ndigen {I}ntegration von
  {K}onstruktionswissen.
\newblock In {\em Design for X--22. DfX-Symposium}, 2011.

\bibitem{Shapes}
O.~Kutz, M.~Bhatt, S.~Borgo, and P.~Santos, eds.
\newblock {\em The Shape of Things, SHAPES 2013}, vol. 1007 of {\em CEUR
  Workshop Proceedings}, 2013.

\bibitem{MossakowskiEtAl13d}
T.~Mossakowski, O.~Kutz, M.~Codescu, and C.~Lange.
\newblock The distributed ontology, modeling and specification language.
\newblock In {\em Workshop on Modular Ontologies, WoMo 2013}, vol. 1081 of {\em
  CEUR Workshop Proceedings}, 2013.

\bibitem{MossakowskiEA13}
T.~Mossakowski, C.~Lange, and O.~Kutz.
\newblock Three semantics for the core of the distributed ontology language
  (extended abstract).
\newblock In {\em International Joint Conference on Artificial Intelligence,
  IJCAI 2013}. IJCAI/AAAI, 2013.

\bibitem{MossakowskiEA07}
T.~Mossakowski, C.~Maeder, and K.~L{\"u}ttich.
\newblock The {H}eterogeneous {T}ool {S}et, {\textsc{hets}}.
\newblock In {\em Tools and Algorithms for the Construction and Analysis of
  Systems, TACAS 2007}, vol. 4424 of {\em LNCS}, pp. 519--522. Springer, 2007.

\bibitem{PahBei:ed07}
G.~Pahl, W.~Beitz, J.~Feldhusen, and K.-H. Grote.
\newblock {\em Engineering Design}.
\newblock Springer Verlag, 3rd edition, 2007.

\bibitem{RabKoh:WSMSML13}
F.~Rabe and M.~Kohlhase.
\newblock A scalable module system.
\newblock {\em Information \& Computation}, 0(230):1--54, 2013.

\bibitem{Roth.1994}
K.~Roth.
\newblock {\em {K}onstruieren mit Konstruktionskatalogen}.
\newblock Springer, Berlin, 1994.

\bibitem{pellet}
E.~Sirin, B.~Parsia, B.~C. Grau, A.~Kalyanpur, and Y.~Katz.
\newblock Pellet: A practical {OWL-DL} reasoner.
\newblock {\em Web Semantics: Science, Services and Agents on the World Wide
  Web}, 5(2), 2007.

\bibitem{VDI2218}
VDI.
\newblock {\em Informationsverarbeitung in der Produktentwicklung --
  Feature-Technologie -- VDI 2218}, 1999.
\newblock {\emph{(Information technology in product development -- Feature
  Technology)}}.

\bibitem{VDI2221}
VDI-Gesellschaft Entwicklung Konstruktion Vertrieb.
\newblock {\em Methodik zum Entwickeln und Konstruieren technischer Systeme und
  Produkte}, 1995.
\newblock English title: Systematic approach to the development and design of
  technical systems and products.

\end{thebibliography}
a\providecommand\seen{seen } \providecommand\selfedit{}
  \providecommand\webpageat{web page at } \providecommand\homepageat{home page
  at } \providecommand\projectpageat{project page at }
  \providecommand\systempageat{system home page at }
  \providecommand\svnrepoat{Subversion repository at }
  \providecommand\January{January} \providecommand\February{February}
  \providecommand\Feb{February} \providecommand\March{March}
  \providecommand\April{April} \providecommand\May{May}
  \providecommand\June{June} \providecommand\July{July}
  \providecommand\August{August} \providecommand\September{September}
  \providecommand\October{October} \providecommand\November{November}
  \providecommand\December{December} \providecommand\AUSTRALIA{Australia}
  \providecommand\ROMANIA{Romania} \providecommand\MEXICO{Mexico}
  \providecommand\ITALY{Italy} \providecommand\USA{USA}
  \providecommand\IRELAND{Ireland} \providecommand\HUNGARY{Hungary}
  \providecommand\JAPAN{Japan} \providecommand\CANADA{Canada}
  \providecommand\SPAIN{Spain} \providecommand\NETHERLANDS{Netherlands}
  \providecommand\UK{UK} \providecommand\SWEDEN{Sweden}
  \providecommand\GERMANY{Germany} \providecommand\openmath{OpenMath}
  \providecommand\fc{forthcoming} \providecommand\PROC{Proceedings}
  \providecommand\omdoc{OMDoc} \providecommand\activemath{ActiveMath}
  \hyphenation{Wiki-Sym}

\end{document}